\documentclass[aps,nofootinbib,twocolumn,superscriptaddress]{revtex4}
\usepackage{lmodern}
\usepackage[T1]{fontenc}
\usepackage{textcomp}
\usepackage{epsf,latexsym}
\usepackage{amsfonts}
\usepackage{graphicx}
\usepackage{color}

\begin{document}

\title{Testing the nonclassicality of gravity
with the field of a single delocalized mass}
\author{Alessandro Pesci\footnotetext{e-mail: pesci@bo.infn.it}}
\affiliation
{
INFN, Sezione di Bologna, Via Irnerio 46, 40126 Bologna, Italy}
\author{Pierbiagio Pieri\footnotetext{e-mail: pierbiagio.pieri@unibo.it}}
\affiliation
{
Dipartimento di Fisica e Astronomia, Universit\`a di Bologna, 
Via Irnerio 46, 40126 Bologna, Italy}
\affiliation
{
INFN, Sezione di Bologna, Via Irnerio 46, 40126 Bologna, Italy}

\begin{abstract}
Most of the existing proposals for laboratory tests of a 
quantum nature of gravity  are based on the use of two delocalized masses 
or harmonically bound masses  prepared in pure quantum states with large enough spatial extent. 
Here a setup is proposed that is based on a single delocalized mass 
coupled to a harmonically trapped test mass (undergoing first expansion
and then compression) that moves under the action of gravity.
We investigate the in-principle feasibility of such an experiment,
which turns out to crucially depend on the ability to tame
Casimir-Polder forces.
We  thus proceed with a design aimed at achieving this,
trying at the same time to take advantage of these forces
rather than only fighting them.

\end{abstract}

%\pacs{}

\maketitle

%%%%%%%%%%%%%%%%%%%%%%%%%%%%%%%%%%%%%%%%%%%%%%%%%%%%%%%%%%%%%
\section{Introduction}

No direct proof exists so far of the quantum nature 
of the gravitational field.
The smallness of the Planck length $l_P$,
at whose scale the quantum aspects of gravity 
should unavoidably appear, 
has always been a formidable obstacle to any attempt
to check for such quantum features at laboratory scales.
This might be about to change, however,
due to the prodigious progress accumulated  over the years
in sensing and controlling quantum systems,
as well as to some new twists \cite{Carney, Huggett}.

The basic idea, which dates back at least to 
an observation by Feynman \cite{Feynman},
is to look at the gravitational field  
sourced by a quantum system in a superposition of states,
prototypically corresponding to spatially separated states.
The effects on a test mass at a distance from the source should depend on 
the nature of the field.
If the gravitational field is quantum,
the test mass should experience a superposition
of field states, each state leading to a
different time evolution of the test mass,  with 
a building up of entanglement between test and source masses during 
this evolution.
If the field is classical, it is single-valued
and no entangling with the source mass is possible.
%}
This is, e.g., the case of gravity in the semiclassical
approximation \cite{Moller, Rosenfeld}, in which the source 
is quantum and the field 
is classical, with the field experienced by the test mass corresponding 
to the expectation value
of the energy-momentum tensor over the quantum state of the source.

Two contrasting requirements apply in general to quantum gravity experiments.
On one hand, large masses are desired to produce 
large gravitational fields and thus amplify the signal
to be measured. On the other hand, the larger the mass is, the more difficult
it is to control  quantum decoherence by the environment \cite{Aspelmeyer}.   
As a matter of fact,
there is still a huge gap \cite{Sidajaya} between 
the largest masses for which an exquisite quantum control
of the position  has been attained (approximately equal to 
$10^{-16}$ kg \cite{Weiss})
and the smallest masses whose gravitational fields
have been directly measured (approximately equal to 
$10^{-4}$ kg \cite{Westphal}).
As a consequence, even just revealing the quantum nature
of the source by the classical gravitational field it produces,
as in the semiclassical approximation,
 is  still something far in the future 
 (see \cite{Sidajaya} for an explicit proposal in this respect).  

In view of this situation, 
many configurations have been proposed
that allow for an increase in the sensitivity 
to possible quantum 
gravitational effects \cite{Lindner, Derakhshani, Carlesso1, Carlesso2}, 
invariably involving, however, masses 
somewhat larger than what is granted by
present limitations on quantum control.
Recently, 
a very promising new twist has been given
by methods in which
the information on the effects of the gravitational field
is encoded
in the variations of the quantum mechanical phase of two 
quantum systems coupled only by gravity, 
each system being in a superposition of states
(an early description of these methods was given in \cite{Bose_yt} and subsequently published in \cite{Bose}).
These ideas have been applied to pairs of 
delocalized particles, as in \cite{Bose, Marletto},
or to pairs of quantum harmonic oscillators, 
as in  \cite{Al Balushi, Krisnanda} and \cite{Weiss}.)
The evolution of the quantum phase, 
driven by the gravitational field sourced by the masses,  
would lead to the creation of detectable entanglement
if the field is treated as quantum.
This is an example of a general strategy for any 
realistic attempt at evidencing the quantum nature of gravity 
in a laboratory:
Focus on verifying features that cannot be explained 
in a classical setting rather than
searching for $O(\hbar)$ corrections to classical results
 (which would be invariably too small, given the smallness 
of the Planck length $l_P$) \cite{Marshman}.
In the  above proposals, masses greater than approximately 
$10^{-15}-10^{-14} \, \rm kg$ have been
considered.

In addition, methods that do not require the occurrence of 
entanglement have been proposed.
 One example is based on the detection of non-Gaussianity in
 Bose-Einstein condensates \cite{Howl}.
 A second recent proposal is instead based on 
 the measurement of a properly defined 
 classical simulation fidelity
 between the actual state of a gravitationally interacting  system 
 undergoing a time evolution and the expected
 time-evolved state if the gravitational field were quantum \cite{Lami}.
 
A difficulty in most of these methods
\cite{Bose,Marletto,Al Balushi,Krisnanda,Weiss} 
is the need to have complete control of
the evolution of two spatially delocalized
or spatially extended wavepackets
against environmental decoherence \cite{Aspelmeyer}.
Furthermore, a general issue in essentially all 
proposals \cite{Bose,Marletto,Al Balushi,Krisnanda,Weiss,Lami} 
is the presence of Casimir-Polder 
forces \cite{Casimir1, Casimir2, Power} which become dominant  
over gravitational effects
at small distances, severely constraining how close to the source one can probe the field, 
thus limiting the intensity of the gravitational effects one is looking for.
Taking advantage of some recent ideas  \cite{Westphal, van de Kamp, Yi}
on how to attenuate Casimir-Polder effects,
the aim of the present paper is to reconsider the case
of a single delocalized particle and propose 
a configuration that seems in principle able
to discriminate between a classical and a 
quantum description of the gravitational field.

%%%%%%%%%%%%%%%%%%%%%%%%%%%%%%%%%%%%%%%%%%%%%

\section{Entangling gravitational field}

\begin{figure}[t]
  \includegraphics[width = 8.6 cm]{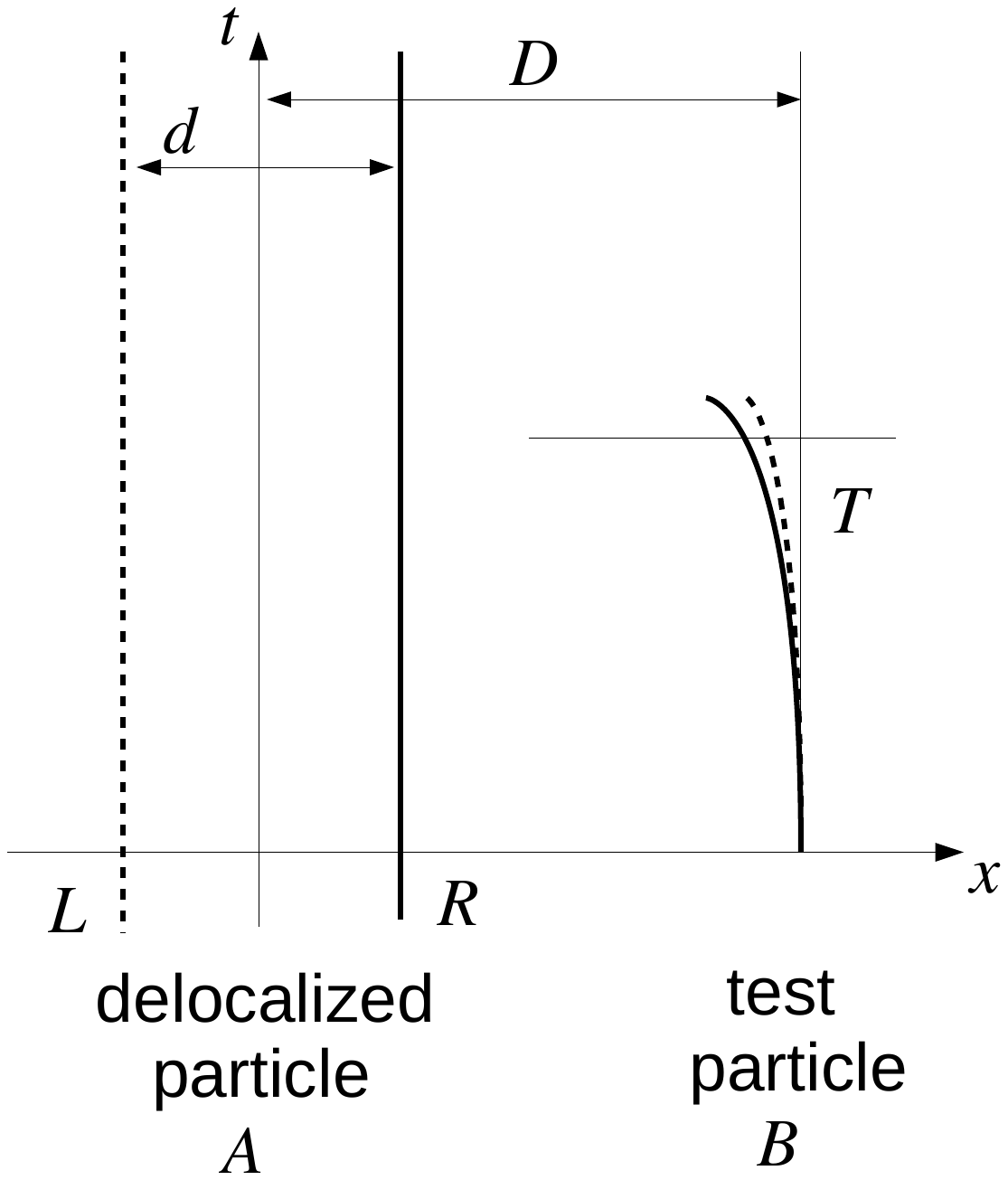}
  \caption{Basic conceptual scheme for the investigation of
the quantum nature of the gravitational field.
A test particle $B$ probes the field sourced by a particle 
$A$ that is in a state delocalized between $L$ and $R$,  assuming the field is entangled with the superposed locations.}
  \label{fig1}
\end{figure}

The basic idea behind 
the investigation of
the quantum nature of the gravitational field 
can be illustrated as follows.
A particle $A$, with mass $m_A$, is prepared 
with its center of mass in a superposition of states 
spatially separated by a distance $d$.
The test particle $B$ with mass $m_B$ 
 is placed at a distance $D$ from $A$ (see Fig.~\ref{fig1}).   
At time $t = 0$ the trapping potential that keeps the particle $B$ 
localized is switched off
so that $B$ is free to move
in the field sourced by $A$.
After a time $T$, the position of $B$ is measured. The outcome will
be different if the gravitational field is quantum or classical.

If the field is quantum, at $t = 0$ the system $A \Phi B$,
formed by the two particles $A$ and $B$  and the field $\Phi$
sourced by $A$, is in a product state
between subsystems $A\Phi$ and $B$. 
It is, however, entangled within the subsystem $A\Phi$:
\begin{eqnarray}\label{zeno46.2}
|\psi(0)\rangle = 
\frac{1}{\sqrt{2}} \,\,
\big(|x_L^A\rangle |\phi_L\rangle + |x_R^A\rangle |\phi_R\rangle\big) 
\,\, |x^B(0)\rangle
\end{eqnarray}
(here the subscripts label the branch that sources the field).

Assuming no effects from the environment,
at time $t$ the two branches evolve to
\begin{eqnarray}\label{zeno46.3}
|x_L^A\rangle |\phi_L\rangle |x^B(0)\rangle 
&\rightarrow& |x_L^A\rangle |\phi_L\rangle |x_{\phi_L}^B(t)\rangle
\nonumber \\
|x_R^A\rangle |\phi_R\rangle |x^B(0)\rangle
&\rightarrow& |x_R^A\rangle |\phi_R\rangle |x_{\phi_R}^B(t)\rangle 
\end{eqnarray}
with $x_{\phi_L}^B(t)$ and $x_{\phi_R}^B(t)$
the evolution of $B$ in the fields $\phi_L$  and $\phi_R$, respectively.
One thus has
\begin{eqnarray}\label{zeno46.4}
|\psi(t)\rangle = \frac{1}{\sqrt{2}} \,
\left[|x_L^A\rangle |\phi_L\rangle |x_{\phi_L}^B(t)\rangle +
|x_R^A\rangle |\phi_R\rangle |x_{\phi_R}^B(t)\rangle\right]
\end{eqnarray}
with entanglement between $A\Phi$ and $B$
and with 
$|x_{\phi_L}^B(0)\rangle =  |x_{\phi_R}^B(0)\rangle = |x^B(0)\rangle$.
After a certain time $T$, the states $|x_{\phi_L}^B(t)\rangle$ and  
$|x_{\phi_R}^B(t)\rangle$ will correspond to wave packets sufficiently
apart in space to be in practice orthogonal to each other.

If, on the contrary, the field is classical 
and is sourced by the expectation value
$\langle T_{ab}\rangle$ of the energy-momentum tensor,
at $t=0$ one has
\begin{eqnarray}\label{zeno46.5}
|\psi(0)\rangle = 
\frac{1}{\sqrt{2}} \, \big(|x_L^A\rangle + |x_R^A\rangle\big) \, |x^B(0)\rangle.
\end{eqnarray} 
Now $B$ is under the influence of the field 
$\phi = \phi(\langle T_{ab}\rangle)$,
yielding
\begin{eqnarray}\label{zeno46.7}
|\psi(t)\rangle = 
\frac{1}{\sqrt{2}} \, \big(|x_L^A\rangle + |x_R^A\rangle\big) 
\, |x^B(t)\rangle,
\end{eqnarray}
with $B$ not discriminating the branch of $A$,
thus remaining unentangled for all times $t$.

\begin{figure}[t!]
  \includegraphics[width = 8.6 cm]{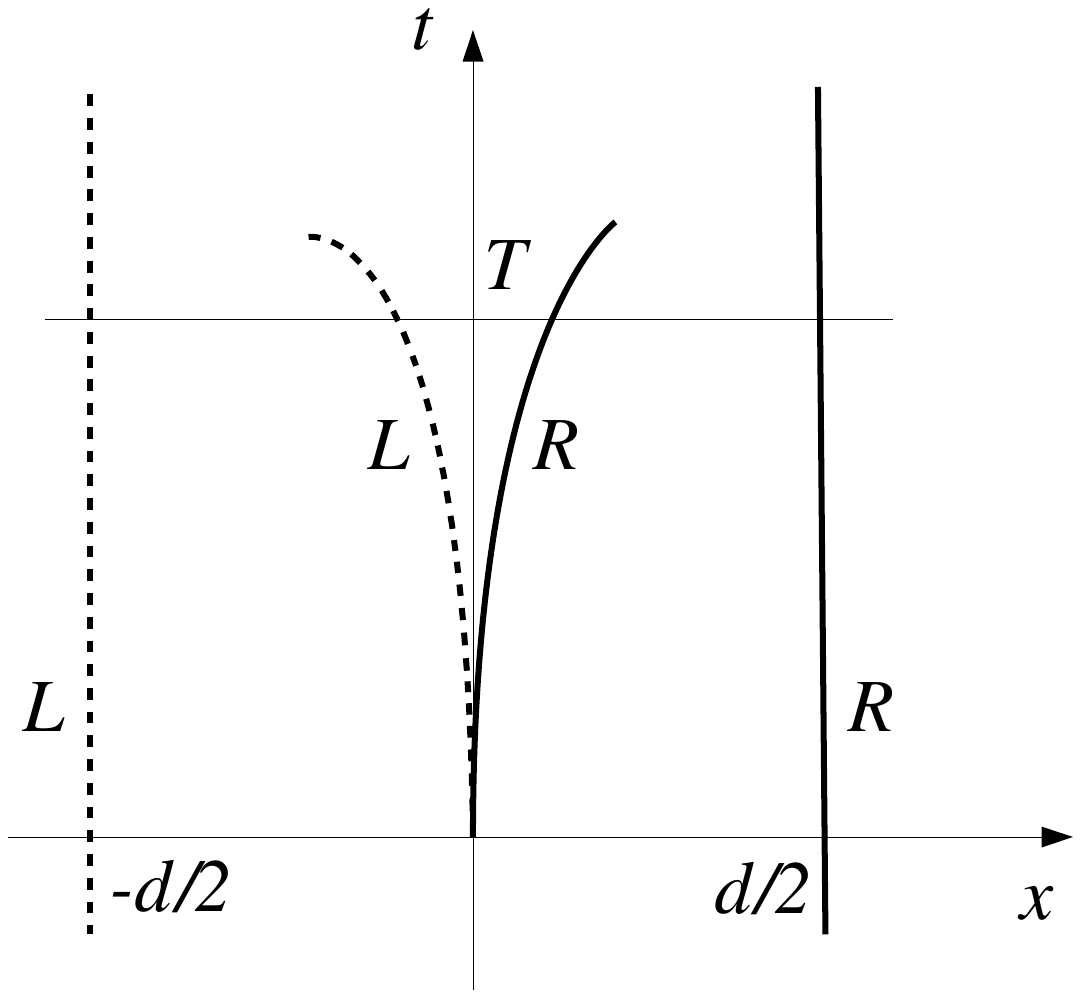}
  \caption{Same as in Fig.~1 but with an improved configuration
to enhance the displacement of the test particle.}
  \label{fig2}
\end{figure}

It is crucial to have some flavor of the displacements
one can have for the test particle
in the two branches.
An informal estimate can be obtained as follows. 
First of all, we note that the configuration displayed in Fig. \ref{fig1} 
is not optimal since the displacements at time $T$ of $B$ from 
the initial position $x^B(0)$ are in the same direction
(towards $A$) for both $L$ and $R$ branches (and in the classical field case as well).
The separation $\Delta x$ between the displacements 
for the two branches is thus
proportional to the absolute value of the difference of the 
(nearly equal and weak) fields in the two branches.
A much better configuration is obtained by placing $B$ {\it between}
 the superposed positions as in Fig.~\ref{fig2};
in this case $\Delta x$ is proportional to the sum
of the absolute values of the fields.

As already mentioned, one has to fight against Casimir-Polder forces
(more on this later).
A distance $l_{\rm CP} \approx 160 \, {\rm \mu m}$ guarantees 
(for silica masses, independently of their size) that 
the ratio $E_{\rm CP}/E_G < 0.1$, where $E_{\rm CP}$ and  $E_G$
are the Casimir-Polder and gravity potential energies, respectively. 
Choosing  $d/2 = l_{\rm CP}$,  one gets
$
\tilde x \equiv \frac{\Delta x}{2} =
\frac{1}{2} a_x T^2 = 2 \frac{G m_A}{d^2} T^2, 
$
where $a_x$ is the acceleration of $B$ and
$G$ is Newton's constant.
For the (quite optimistic) values  $m_A = 10^{-11} \, {\rm kg}$ 
and $T = 1 \, {\rm s}$ one obtains
$
\tilde x = 1.3 \times 10^{-14} \, {\rm m}.
$

Comparing this value of $\tilde x $ with the spread $\sigma_x$ achievable
for masses cooled to the ground state
in an optical tweezer \cite{Romero-Isart, Tebbenjohanns, Delic},
$\sigma_x \approx 0.1-1 \,\, {\rm pm}$ 
\cite{Weiss},
one sees that we are far from the required sensitivity.
This is mainly due to the large distances involved 
to reduce the importance of Casimir-Polder forces
(for the assumed mass and evolution time).
This is in spite of having been very generous 
with our assumptions on technical capabilities.  
We have indeed taken quite a large value for the delocalized
mass and the most convenient geometric configuration
 to enhance the displacement (this resulting, however,
in uncomfortably large separations 
between the two states of the delocalized mass; cf.~\cite{Aspelmeyer}).
We have also assumed that the state superposition can be kept
even with the test particle placed between $L$ and $R$
and for an evolution time $T$
which is very long compared to typical decoherence 
times (and which requires a high control of the displacements 
due to free fall).

This indicates that in order to have any chance 
to probe the quantum nature of gravity 
along the above lines, one should first 
find a way to overcome the limitations
imposed by Casimir-Polder interactions.
A description of how this might be attempted
is the aim of the following sections.

%%%%%%%%%%%%%%%%%%%%%%%%%%%%%%%

\section{Taming (and exploiting) the Casimir--Polder forces}

The aim of this section is to describe ways to circumvent
the issues posed by Casimir-Polder forces.
We will see that setups can be considered
not only in which this looks possible,
 but which also allow one to benefit
 from these forces rather than 
 simply  fighting them 
(somewhat similarly to Ref.~\cite{Pedernales},
which contemplated the possibility to use the Casimir force to
strongly couple an ancillary system to gravitationally interacting 
masses).

The first step in this direction is the observation
that Casimir-Polder  forces can be conveniently screened
by using conducting plates.
This was indeed used in \cite{Westphal}
to allow masses to be close enough to each other
to make detection of  their gravitational field possible.
Screening with conducting plates was also proposed 
in \cite{van de Kamp, Yi} 
as a modification of the proposal \cite{Bose, Marletto}
aiming at revealing gravitational entanglement 
between two delocalized particles close to each other.
Indeed, the insertion of a perfectly conducting plate
between the delocalized particles allows one 
to screen the Casimir-Polder
force between the particles, replacing it with the force 
between each particle and the plate,
of the same nature but of lower strength.
This allows one to reach smaller distances between particles
(roughly from the $160 \, \mu{\rm m}$ mentioned before to
approximately $40-50 \, \mu{\rm m}$ \cite{van de Kamp, Yi}),
thus enhancing the gravitational effects,
and allowing for milder requirements for the required separation $d$
between the superposed positions.  

In the setup of Fig.~\ref{fig1} one could benefit from this screening
by inserting a conducting plate between
$A$ and $B$.
However,  in view of the drawbacks of this  
configuration discussed above, only
a marginal gain could be reached in this case.

It is not immediately clear how to use this screening 
for the more promising setup of Fig.~\ref{fig2}.
Indeed, it looks that one has to introduce
two conducting plates, placed symmetrically on the two sides of the test
particle in order to separate it from the two possible positions of the 
delocalized source particle.
This is clearly a problem, since it is not at all obvious how 
a delocalized state could be prepared in the presence of
conducting plates between the two superposed positions.  

\begin{figure}[h!]
  \includegraphics[width = 8.6 cm]{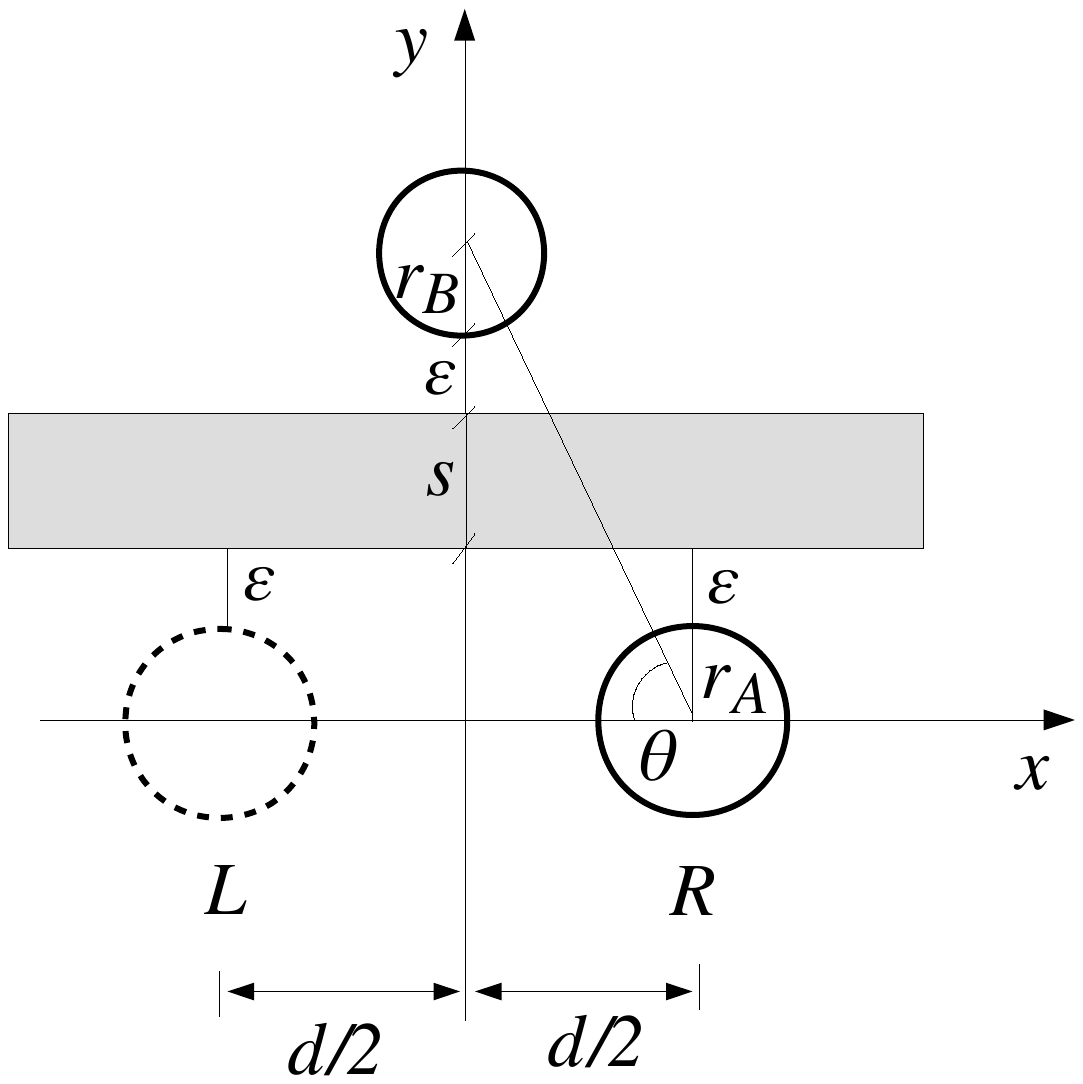}
  \caption{Configuration with a horizontal metallic plate shielding
the Casimir-Polder forces between test and source masses,
helping, at the same time, the levitation of the (heavier) delocalized particle $A$.}
  \label{fig3}
\end{figure}

A more convenient arrangement is shown in Fig.~\ref{fig3}.
In it, a single conducting plate screens the Casimir-Polder forces 
between the two particles. In addition, the 
Casimir-Polder force between the plate and 
the delocalized particle acts in the same way 
for the two superposed positions. In this way, 
the coherence of the delocalized state is not affected.
Finally, the Casimir-Polder force by the plate
on the test particle is in the direction orthogonal to the plate 
and thus does not produce displacements 
parallel to the plate, which are the ones relevant to detect
the gravitational effects we are after.
The same considerations apply to the gravitational forces
between test and source masses and the plate.
As a matter of fact,
%imperfect parallelism of the plate,
irregularities on the surfaces such as random patch potentials and density fluctuations in the materials
could possibly affect both the coherence of the delocalized state
and the displacements of the test particle. 
We will return to this issue later on, when focusing on Casimir-Polder
effects.

Taking the $x$ axis in the direction connecting the two
superposed positions $L$ and $R$, the displacement $\tilde x$
induced along $x$ by $A$ in branch $|x_R\rangle$
can be estimated as
$\tilde{x}= \frac{1}{2} a_x T^2$, with
\begin{eqnarray}
a_x &=&
\frac{G \, m_A}{(d/2)^2 + y_B^2} \, \cos\theta\\
&=&
\frac{G \, m_A}{(d/2)^2 + y_B^2} \,\, \frac{d/2}{\left[(d/2)^2 + y_B^2\right]^{1/2}}\\
&=&
\frac{2}{3 \sqrt{3}} \, \frac{G \, m_A}{y_B^2},
\label{zeno48.1}
\end{eqnarray}
where $y_B = r_A + 2 \epsilon + s + r_B$
(see Fig.~\ref{fig3}).
To get Eq.~(\ref{zeno48.1}) we used the most favorable 
geometric conditions for given $y_B$,
namely the value of $d/2$ that gives the largest $x$ component $F_x$ 
of the gravitational force
between $A$ and $B$, corresponding to $d/2 = y_B/\sqrt{2}$.

By taking $T = 1 \, {\rm s}$,
$m_A = 10^{-11} \, {\rm kg}$, $m_B = 10^{-14} \, {\rm kg}$ 
(corresponding to $r_A = 9.7 \, \mu{\rm m}$, and $r_B = 0.97 \, \mu{\rm m}$,
respectively,  for silica),
$\epsilon = 1 \, \mu{\rm m}$, and 
$s = 1 \, \mu{\rm m}$,
the optimal separation turns out to be just $d \approx 2 \, r_A = 19.4 \, \mu{\rm m}$,
with a resulting value for
$
\tilde x \approx 0.7 \, {\rm pm}
$
that is right within the range we mentioned above
as reasonably reachable with optical tweezers.
For $m_A < 10^{-11} \, {\rm kg}$, the optimal separation $d$
is within $2 \, r_A < d < 19.4 \, \mu{\rm m}$.

In the above estimate we have considered
 $m_B = 10^{-14} \, {\rm kg}$.
This value is about 50 times 
the heaviest masses that 
at present can be harmonically controlled and cooled to the  
ground state
(which are made of about $10^{11}$ nucleons
\cite{Delic, Tebbenjohanns2},
corresponding to approximately $2 \times 10^{-16} \, {\rm kg}$). 
The spread of $x_B$ in the ground state 
is
\begin{eqnarray}\label{zeno18.6}
\sigma_x = \sqrt{\frac{\hbar}{2 \, m_B \omega}},
\end{eqnarray}
with $\omega$ ($\sim 2 \pi \times 100 \, {\rm k Hz}$ 
\cite{Delic, Tebbenjohanns2}) 
the angular frequency of the trapping potential.
For $m_B = 10^{-14} \, {\rm kg}$, Eq.~(\ref{zeno18.6}) gives
$\sigma_x = 0.09 \, {\rm pm}$,
and the $\tilde x$ we have just obtained
appears definitely detectable.

Here
we are assuming that the spread $\sigma_x(T)$ 
in the position of $B$ after a time $T$ is the same
as the initial one, which is determined by Eq.~(\ref{zeno18.6}).
One might wonder if this assumption is justified.
Indeed, the initial momentum spread $\sigma_p(0)$
implies that after a time $t$  one has
$
\sigma_x^2(t) = \sigma_x^2(0) + \sigma_p^2(0) \,
\frac{t^2}{m^2}
$
(see e.g. \cite{Gasio}),
with $\sigma_p(0) \geq \frac{\hbar}{2 \, \sigma_x(0)}$.
For $t > 0$, this implies
$\sigma_x^2(t) > \hbar t/m$
\cite{Braginski} 
(corresponding to the standard quantum limit, 
see e.g. \cite{Joos}). 
Using $t = 1 \, {\rm s}$ and
$m = 10^{-14} \, {\rm kg}$,
one obtains
$\sigma_x(1 \, {\rm s}) > 10^{-11} \, {\rm m}$,
which is about
two orders of magnitude larger that the displacement $\tilde x$
one wishes to measure.
This apparently implies that our assumption is not justified.

The  spread $\sigma_x$ of the test particle's wave function could 
however be kept under control by using the 
loop protocol proposed in \cite{Weiss}.
Specifically, through an appropriate alternation between 
an inverted harmonic potential and a harmonic one,  
the spatial probability distribution 
of a levitated nanoparticle is first expanded to scales
much larger than the initial one and then converted back 
to the original configuration.
Such a loop protocol works also in the presence 
of an external constant force $F$ \cite{Weiss},
like gravity (along $x$) in our case. In this way,
the wave function of the test particle at  $t = T$ can have the same spread
as at $t=0$, while being displaced by the gravitational force $F$, as 
required to measure the effect we are after. 
We thus assume that this or a similar protocol
ensuring $\sigma_x(T) = \sigma_x(0)$
is applied to the test particle $B$. 
This assumption is still quite optimistic at
present, since the value $m_B = 10^{-14} \, {\rm kg}$ 
we are assuming for the test mass is two
orders of magnitude larger than the masses for which
full quantum control has been achieved so far. 
We are however confident that progress will 
grant full quantum control for masses as large as
 $10^{-14} \, {\rm kg}$ in the near future.
 
It should be noted, however, that ground-state cooling
of the test mass $m_B$ and preserving its spread
during the experiment are not the main difficulties
to be addressed in a practical implementation of our proposal.
A major challenge is the preparation of a delocalized state
for the required large values of the source mass $m_A$ (of the order of $10^{-14}$ kg, at least)
and separations $d$ between the superposed locations (of the order of $10 \mu$m). 
The state of the art is that
separations 
of the order of a fraction of a micron   
have been achieved for macromolecules of several 
thousand of atoms, corresponding
to masses of a fraction of $10^{-22} \, {\rm kg}$ \cite{Fein}.
One sees that, for what we are proposing,
experiments are off at present by several orders of magnitude
in the mass and/or separation.
It is however worth mentioning that experimental schemes for the preparation
of masses approximately equal to $10^{-14}$ kg 
in a spatial quantum superposition state 
with an extent of the order of the micron 
have been proposed recently \cite{Pino}.

A further challenge is keeping quantum coherence
for the quite long evolution times we are requiring here
($T \approx 1 \, {\rm s}$).
Coherence times of the order of 1~s have been shown 
for solid-state spin systems cooled down at 77 K \cite{Bar-Gill}
(coherence times become of the order 1 ms  at room temperature \cite{Herbschleb}).  
 For the kind of massive objects we are interested in, coherence times of the order of
 1~s have been considered in the experimental scheme  for the preparation of delocalized states just mentioned \cite{Pino}. 
 On the other hand, 
 Ref.~\cite{Folman}  has recently pointed out that, for massive objects, 
an intrinsic decoherence channel, which adds to decoherence from the environment,  
originates from the phonons within the object, 
which are almost unavoidably excited during 
the creation of the superposed state.
A way to overcome this problem would be to make the applied force used to
create the superposed state sufficiently homogeneous over the scale
of the massive object, in order to minimize the creation of phonons.
It is thus clear that both the preparation of the superposed state with the desired
parameters and maintaining its quantum coherence for the required
long times represent extremely difficult tasks, which will require a technological leap to be 
successfully addressed.
 
 Returning to the discussion of Casimir-Polder effects, we notice that
with the chosen value $\epsilon = 1 \, \mu{\rm m}$ for the distance
between the two masses and the plate, Casimir-Polder forces by the
plate on the two masses will be quite large, much larger, in particular,
than the gravitational interaction
between source and test masses. 
They will however have a vanishing component along $x$ and
thus will not spoil the measurement of the displacement
along $x$ of our interest.
Actually, in the configuration of Fig.~\ref{fig3},
Casimir-Polder forces could even be used to contrast
the effects of earth's gravity on particle $A$ (aiding, or altogether
replacing, optical levitation). 
For particle $B$, a second plate could be arranged above it, 
parallel to the first one at a distance less than $\epsilon$ in order 
to partially
(or completely) balance earth's gravity.
In the case of a good balance of the forces along the vertical
(which might be achieved by adding
also a third plate, 
this time below the delocalized particle $A$,
to regulate the overall Casimir-Polder force on the latter)   
one would not have appreciable displacements $\Delta y$ along $y$, 
even for times as large as $T \sim 1 \, {\rm s}$
(which would lead to $\Delta y \approx 5 \, {\rm m}$ for free fall).
One could thus potentially perform the experiment
in an earth-based laboratory with no need to go after free fall,
use optical levitation, or envisage it as a space experiment.

In order to estimate the Casimir-Polder force $F_{\rm CP}$ between particles
and plates,
we can use its expression when the distance of closest approach
to the plate is $l \leq r_{A,B}$,
\begin{eqnarray}\label{zeno27.2}
F_{\rm CP} = 2 \pi \, r \Big(\frac{1}{3} \, \frac{\pi^2}{240} 
\, \frac{\hbar c}{l^3}\Big),
\end{eqnarray}
where $c$ is the speed of light in vacuum and $r$
is the radius of the spherical masses
 \cite{Lamoreaux}.  
Using this formula, one finds that,
with $l = 0.3 \, \mu{\rm m}$,
$F_{\rm CP}$ alone is able to balance the weight of $B$,
with no need of optical levitation.
Indeed, for a silica spherical mass $m_B = 10^{-14} \, {\rm kg}$, one has
$r_B = 0.97 \, {\mu{\rm m}}$, 
yielding  $F_{\rm CP} = 0.98 \times 10^{-13} \, {\rm N} \approx m_B \, g$,
where $g$ is earth's gravity.
  
In comparison, 
the gravitational force $F_{G-AB}$ between $A$ and $B$
is extremely feeble.
From (\ref{zeno48.1}), with the same conditions discussed above
(in particular, $m_A = 10^{-11} \, {\rm kg}$ and $m_B = 10^{-14} \, {\rm kg}$),
one gets 
$F_{G-AB} \approx 2.4 \times 10^{-26} \, N$,
which is about 13 orders of magnitude smaller than the Casimir-Polder
forces.
Such a huge difference makes it clear that,
for the actual feasibility of the experiment,  
it is crucial  to reach an extreme accuracy in the preparation of the setup.
In particular, imperfections  in the planarity
of the metallic plate or in the spherical shape of 
the particles
could produce side effects 
in the $x$ direction that, even if much smaller than $F_{\rm CP}$, 
could possibly overwhelm $F_{G-AB x}$ or lead to
variations of $F_{\rm CP}$ in the two superposed locations
of the source particle that could spoil its coherence.
Effects of this kind,
possibly also changing with time, 
could originate from patch potentials and density fluctuations,
as well as from any kind of irregularities of the surfaces.

Considering that our aim is to exploit Casimir-Polder forces
rather than finding ways to suppress them,
it is important to analyze this kind of effect,
which could hinder the feasibility of the proposed experiment. 
For the source mass, the main issue could be
a variation of  the Casimir-Polder force between the mass and the plate in the two superposed locations due to the above 
random fluctuations. Previous studies \cite{Klimchitskaya-1996,Klimchitskaya-2009} have shown that, for a sphere with
radius $R$ at an (average) distance $l$ from the plate, 
with random corrugations of amplitude $\Delta z \ll l$  for
both surfaces, the Casimir-Polder force does not depend on the position on the plate provided 
 the characteristic lateral scales $\Lambda_p$ and $\Lambda_s$  for 
the surface roughness of the plate and the sphere, respectively, are small compared to 
$\sqrt{R l}$ (for $l \ll R$) or to $l$ 
(for $R \ll l$; see \cite{Bezerra-1999}).
With our numbers for the source mass ($R\approx 10 \mu$m and $l \gtrsim 0.1 \mu$m), 
this would require keeping the amplitude  $\Delta z$ of the surface roughness within 10 nm 
and the lateral scales $\Lambda_s$ and $\Lambda_p$ within 100 nm. 
Under these conditions, the effect of the surface roughness would be just a correction by a multiplicative factor \cite{Klimchitskaya-2009} 
of the expression~(\ref{zeno27.2})  for the force $F_{\rm CP}$, 
with no variations  between the two superposed
positions and thus no spoiling of the delocalized state of the source mass.

Concerning electrostatic patch potentials, which could produce an electrostatic force on top of the Casimir-Polder force, 
a recent work \cite{Garrett} has shown that by applying an 
ion-blocking layer on the surfaces of the sphere and plate, the force produced by
patches can be reduced to $10^{-4} F_{\rm CP}$ for distances between the plate and the sphere of the same order as considered here. 
We speculate that such variations of the force, which could lead to an imperfect balance between the weight of the source mass and the 
Casimir-Polder force by the plate
 on top of it (see in particular the final experimental configuration described in the next section), could be eliminated by an optical levitation
 system with an effective potential conveniently shaped
 to compensate for the residual  force difference (of order $10^{-4}$ the weight of the particle) between the two superposed locations.    

For the test mass, the main issue could be the occurrence
of  lateral Casimir-Polder forces due to irregularities of 
the surface, which could produce a lateral displacement on top of the gravitiational one.
It should be pointed out, however, that for the same random short-scale irregularities just discussed, 
these forces cancel out altogether~\cite{Klimchitskaya-2009}. As a matter of fact,
only  parallel uniaxial sinusoidal corrugations  of the surfaces with the same period 
can produce  nonzero lateral Casimir-Polder forces,  
as  also shown in the seminal experiment \cite{Chen}.
 
We notice here however that,
even in the absence of such irregularities,
the strong Casimir-Polder forces
 $F_{\rm CP}$ by the masses on the plate 
 will likely deform it, with ensuing side effects possibly
larger than those just mentioned. We thus focus on
discussing these effects, which are present even 
for an ideal realization of the experimental setup.

Looking at Fig.~\ref{fig3},
one can see that a deformation of the plate 
produced by $F_{\rm CP}$
when $A$ is in one of the two branches,
for example when $A$ is in $R$,
gives an effective distance between $B$ and the plate
larger to the right than to the left, and thus
a Casimir-Polder attraction stronger towards
the left than towards the right.
One would thus have a Casimir-Polder acceleration along $x$,
induced by the deformation, 
that is opposite
to the one expected from gravitational interaction 
between $A$ and $B$, possibly overwhelming it
and thus  undermining the experiment. 

The deformation induced by a force $F_{\rm CP}$
acting at the center of a square plate of side $\ell$
and thickness $s$,
with clamps at the ends at distance $\ell$, 
say, along $x$, 
can be estimated as \cite{Landau} (see also \cite{van de Kamp}) 
\begin{eqnarray}\label{zeno40.1}
\delta_{\rm def} = \frac{F_{\rm CP} \, \ell^3}{192 \, E I},
\end{eqnarray}
where $E$ is the Young modulus ($E =137 \, {\rm GPa}$ if the plate
is made of copper) and $I = \frac{\ell}{12} s^3$ is the area moment 
of the plate with respect
to the axis through its center, parallel to the plate 
and orthogonal to the $x$ direction. 
If $F_{\rm CP}$  corresponds to the weight of a particle of mass $m$ (perfect balance), 
one has
\begin{eqnarray}
\delta_{\rm def} = 
mg \, \frac{\ell^2}{16 \, E s^3}.
\end{eqnarray}
For $\ell = 1 \, {\rm mm}$, $s = 1 \, \mu{\rm m}$ and $m = 10^{-14} \, {\rm kg}$ ($m = 10^{-11} \, {\rm kg}$ ) this gives
$\delta_{\rm def} = 0.45 \times 10^{-13} \, {\rm m}$
($\delta_{\rm def} = 0.45 \times 10^{-10} \, {\rm m}$).
Strictly speaking, these deformations
add to the deformation already present 
due the action of earth's gravity on the plate 
(which is symmetric about the plate center).

To estimate the effect of deformations of such an amount,
let us compare $F_{\rm CP}$ corresponding 
to a given distance of closest approach $l \approx 0.1 - 10 \, {\mu{\rm m}}$
with the Casimir-Polder force $F_{\rm CP}^*$
one would get if this distance were larger by, say, $\eta = 10^{-13} \, {\rm m}$,
$l' = l + \eta$.
One can roughly think that the left-right imbalance 
in the total Casimir-Polder force 
induced by the deformation should not be larger 
than the difference $F_{\rm CP} - F_{\rm CP}^*$.
From Eq.~(\ref{zeno27.2})
with $\eta \ll l$,  one gets
\begin{eqnarray}\label{zeno38.3}
F_{\rm CP} - F_{\rm CP}^* = 3 \, \frac{\eta}{l} \, F_{\rm CP}. 
\end{eqnarray}
With $\eta = 10^{-13} \, {\rm m}$, 
and the value $F_{\rm CP} = 0.98 \times 10^{-13} \, {\rm N}$
calculated above (corresponding to $l = 0.3 \, \mu{\rm m}$ 
and $r \approx 1 \, \mu{\rm m}$), one gets
$F_{\rm CP} - F_{\rm CP}^* \approx 3 \times 10^{-20} \, {\rm N}$,
which is six orders of magnitude larger than the gravity force $F_{G-AB}$ 
between $m_A = 10^{-11} \, {\rm kg}$ and $m_B = 10^{-14} \, {\rm kg}$.
 
This shows that  to proceed with our program
one should definitely address 
the issue of the induced deformations, including also the deformation of the plate
under its own weight.
The ideal solution would be keeping the deformations 
(at least the dangerous ones, i.e., those able to induce a left-right
asymmetry along $x$ at the actual position of $B$)
small enough.
In particular, for deformations smaller than the zero-point 
motion $\delta_0$ of the plate considered as a harmonic oscillator,
one expects the oscillator to relax to the ground state 
with no significant effect from the (would be) deformation. 

To estimate the zero-point motion of the center of the plate,
we describe it as a harmonic oscillator
with mass $M$ equal to the mass of the plate,
elastic constant $k = \frac{16 \, E s^3}{\ell^2}$ 
[in view of Eq.~(\ref{zeno40.1}) written as
$\delta_{\rm def} = \frac{F_{\rm CP}}{k}$],  
and angular frequency 
$\omega = \sqrt{\frac{k}{M}}$
\cite{van de Kamp}.
Analogously to Eq.~(\ref{zeno18.6}),
the zero-point motion along $y$ of the center
of the plate can be estimated as
\begin{eqnarray}\label{zeno40.5}
\delta_0 &=&
\sqrt{\frac{\hbar}{2 \, M \omega}}
=
\sqrt{\frac{\hbar}{2 \, \sqrt{M k}}}
=
\frac{1}{2 \sqrt{2} \, s} \sqrt{\frac{\hbar}{\sqrt{\rho E}}},
\end{eqnarray}
where $\rho$ is the density of the material the plate is made of.
For copper ($\rho = 8.96 \times 10^3 \, {\rm kg/m^3}$) and
taking $s = 1 \, \mu{\rm m}$, one gets
$\delta_0 = 0.61 \times 10^{-15} \, {\rm m}$.

Actually, the ground-state oscillations of the plate  
alone might in principle overwhelm the effects we are after.  
Indeed, the zero-point motion deviation $\delta$ of the plate from 
its equilibrium position varies along the distance  
connecting the clamps, being $\delta=\delta_0$ maximal in the middle
and $\delta\approx 0$ at the clamps, thus inducing a gradient  
along $x$ of the Casimir force. An upper bound for the effects of
the possible ground-state motion asymmetry 
is provided by replacing $\eta$
with $\delta_0$  in Eq.~(\ref{zeno38.3}), yielding
$F_{\rm CP} - F_{\rm CP}^* \approx 10^{-22} \, {\rm N}$, 
which is four orders of magnitude larger than the gravity 
force $F_{G-AB}$ calculated above.
This means that the ground-state oscillations 
of the plate alone might in principle overwhelm
the effects we are after.

The aim of the following section is thus to modify the 
setup of Fig.~\ref{fig3} and find a configuration 
that overcomes the problems due to ground-state motion, to
the deformations induced by the 
test and source masses, and to
the deformation of the plate by its own weight.

%%%%%%%%%%%%%%%%%%%%%%%%%%%%%%%

\section{Our proposal}

Our proposal 
is based on adjusting the parameters
and the spatial arrangement of masses and plates,
and on taking advantage 
as much as possible
of symmetry considerations,
in order to minimize adverse effects.

First of all, the Casimir-Polder force  balancing the weight of the delocalized mass $A$
should not act on the plate separating $A$ and $B$ (as in Fig.~\ref{fig3}).
We saw indeed that a deformation-induced left-right imbalance in 
the Casimir-Polder force by the separating plate on $B$ could overwhelm the 
gravitational effects between $A$ and $B$. 

\begin{figure}[t]
  \includegraphics[width = 8.6 cm]{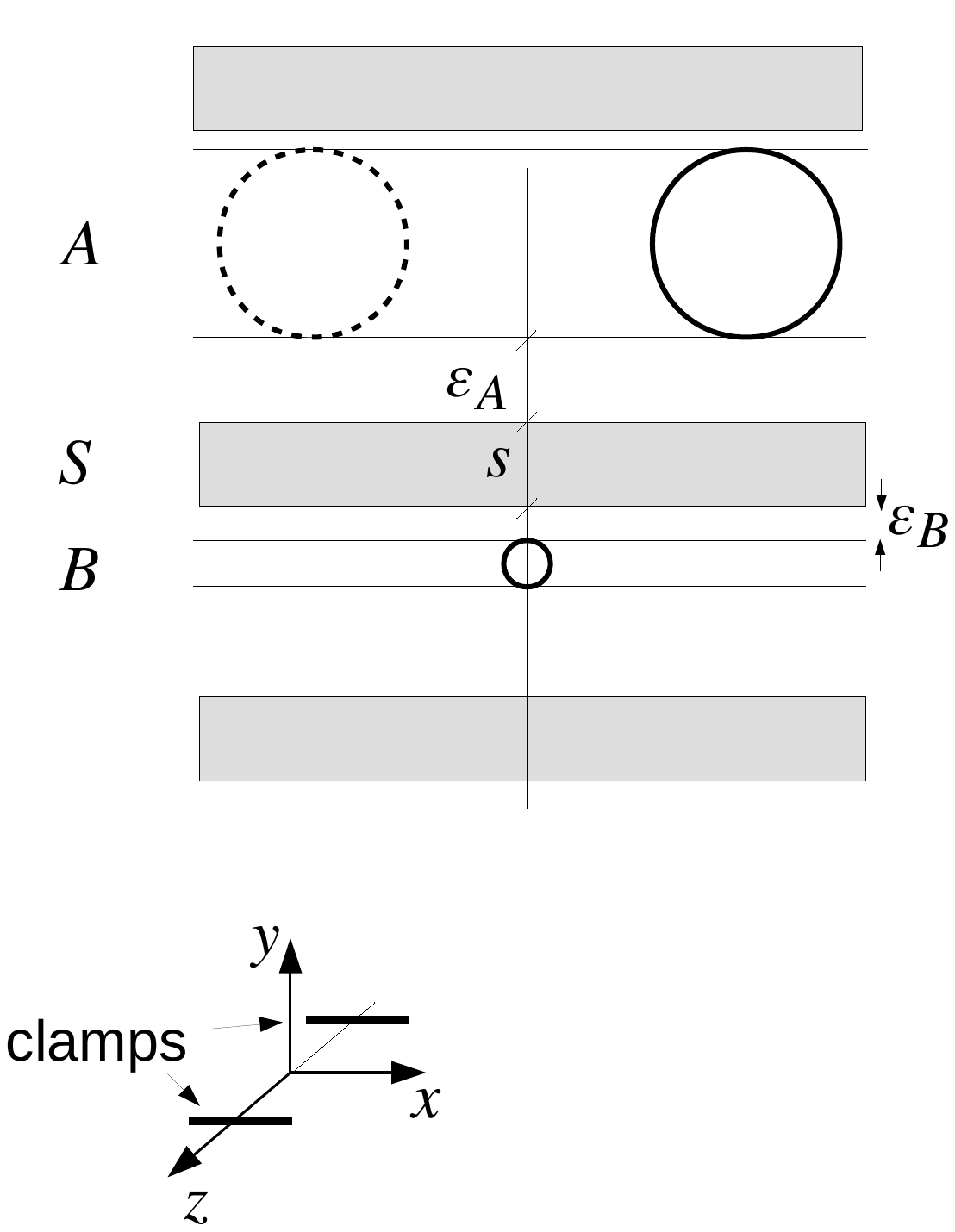}
  \caption{Proposed experimental setup. The metallic plate $S$ screens the
Casimir-Polder forces between the two particles
and partially or fully compensates the weight of $B$
with Casimir-Polder forces.
The additional plate on top of $A$ partially or fully compensates the weight of $A$, while 
the optional plate below $B$ can be used for a fine-tuning of the 
levitation of $B$.
The clamps for the plates are separated by
a distance $\ell$ along $z$.}
  \label{fig4}
\end{figure}

We thus propose to place $A$ above the plate $S$ separating the two masses
and add above $A$ a second plate, as shown in Fig.~\ref{fig4}. 
This second plate, if sufficiently close to $A$, could balance the weight of $A$
through Casimir-Polder forces. At the same time, deformations of this plate 
by the mass $A$ will not affect the superposition
of states of $A$, since the effects of deformations  will be 
identical  in the two branches. 
The separation between $A$ and $S$ can then be chosen
as large as required for keeping the Casimir-Polder deformation
of $S$ below the ground-state oscillation of the plate.    

The test mass $B$ can be placed close to $S$ to compensate its
weight with the Casimir-Polder force of $S$ (aiding, or even completely 
replacing, optical levitation).
The deformation induced by $B$ on $S$, even if somehow larger than 
the ground-state motion of the plate,
is not an issue because it is completely symmetric 
with respect to the position of $B$ 
and thus expected to have no Casimir-Polder effect on the motion of
$B$ along $x$.
One might however possibly expect some higher-order
effects if $B$ is not exactly 
centered with respect to the clamps.

These potential effects, as well as the deformation of the plate $S$ under
its own weight, and the effects due to its ground-state motion can all
be taken to be reasonably under control
if the clamps
on $S$ are put at a distance $\ell$ along the direction $z$ (orthogonal
to $x$, in the plane; see Fig.~\ref{fig4}).
Indeed, with this choice, all of these 
deformations vary
only along $z$ and should  
not produce any effect along $x$.     

A third plate could finally be arranged below $B$ 
(see Fig.~\ref{fig4}) 
to allow for fine adjustments to get an exact balance
of the weight of $B$ (in the absence 
of optical levitation, in particular).

We now proceed to estimate the value of the distance $\epsilon_A$ required
to keep the deformation induced by $A$ below
the ground-state motion of $S$.
To this end, we notice that Eq.~(\ref{zeno40.5}),
for a value $s = 2 \, \mu{\rm m}$ of the thickness of $S$, 
yields the value $\delta_0 \approx 0.3 \times 10^{-15} \, {\rm m}$
for the ground-state motion of the center of the plate
(which is the same for of all points along $x$ 
in the middle of the plate). 
On the other hand, a choice of $\epsilon_A = 3 \, \mu{\rm m}$ yields
for the deformation $\delta_{\rm def}$ (Casimir-Polder) induced by $A$ 
on $S$ the value
$\delta_{\rm def} \approx 0.6 \times 10^{-16} \, {\rm m}$,
as obtained by inserting $l=\epsilon_A $ in Eq.~(\ref{zeno40.1}), with  
 $F_{\rm CP}$ given by Eq.~(\ref{zeno27.2}).  
Here we are considering a square plate 
of copper with side length $\ell = 1 \, {\rm mm}$. 
This value of  $\epsilon_A $ thus guarantees $\delta_{\rm def} \ll \delta_0$,
and no effects are expected from possible deformations induced by $A$.

%%%%%%%%%%%%%%%%%%%%%%%%%%%%%%%

\section{Expected signal}

We can now proceed to estimate the displacements $\tilde x$ of $B$ 
along $x$ produced by the gravitational field sourced by $A$ in each of
the two branches.

\begin{figure}[t]
  \includegraphics[width = 8.6 cm]{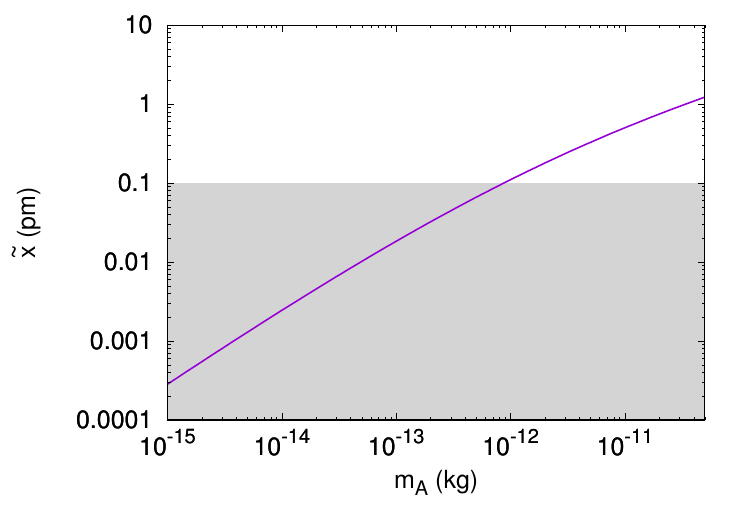}
  \caption{Displacement $\tilde x$ of the test mass $m_B = 10^{-14} \, {\rm kg}$, after a time $T = 1 \, {\rm s}$, 
 as a function of the mass $m_A$ of the delocalized source particle. 
The shaded area shows the displacement below the
 current resolution limit (0.1 pm). Test and source masses are assumed to be made of silica.}
  \label{fig5}
\end{figure}

The displacement can be conveniently written as a function
of the mass $m_A$ of the delocalized particle and of the evolution
time $T$, $\tilde x = \tilde x(m_A, T)$, 
as
\begin{eqnarray}\label{zeno50.1.1}
\tilde x 
&=& \frac{1}{2} \, a_x T^2 \\
\label{zeno50.1.2}
&=&
\frac{1}{3 \sqrt{3}} \, \frac{G m_A}{y_B^2} \, T^2 \\
&=&
\label{zeno50.1.3}
\frac{1}{3 \sqrt{3}} \,
\frac{G m_A}{\left[\left({3\over 4 \pi}\right)^{1\over 3} \left({m_A\over \rho}\right)^{1\over 3} + \tilde y_B\right]^2} \, T^2,
\end{eqnarray}
where
$\rho = 2.6 \times 10^3 \, \frac{\rm kg}{\rm m^3}$
for silica
and
$\tilde y_B \equiv r_B + \epsilon_B + s + \epsilon_A = 6.3 \, \mu{\rm m}$
(for fixed $m_B = 10^{-14} \, {\rm kg}$).
In Eq.~(\ref{zeno50.1.2}) we have assumed that $d$ is optimal 
(namely, the choice that maximizes $a_x$, as described above).
For the above value ${\tilde y_B}\approx 6 \, \mu{\rm m}$, the geometrical constraint $d/2 > r_A$ 
is respected for masses $m_A \lesssim 4 \times 10^{-11} \, {\rm kg}$. 
The distance $d/2$ varies
 from $5.1 \, \mu{\rm m}$ 
to $11.3 \, \mu{\rm m}$ 
as $m_A$ changes from $10^{-14} \, {\rm kg}$ to $10^{-11} \, {\rm kg}$ 
(and $r_A$ varies correspondingly from $0.97 \, \mu{\rm m}$ to $9.7 \, \mu{\rm m}$). 

Figure \ref{fig5}  shows the displacement $\tilde x$ as a function of the mass $m_A$
of the delocalized particle assuming an evolution time $T = 1 \, {\rm s}$ and a mass 
$m_B = 10^{-14} \, {\rm kg}$ for the test particle which, according to (\ref{zeno18.6}),
 gives a spatial resolution better than
0.1 pm.
One sees from Fig.~\ref{fig5} that the displacement 
should in principle be detectable for a source mass
$m_A \gtrsim 10^{-12} \, {\rm kg}$. 

\begin{figure}[t]
  \includegraphics[width = 8.6 cm]{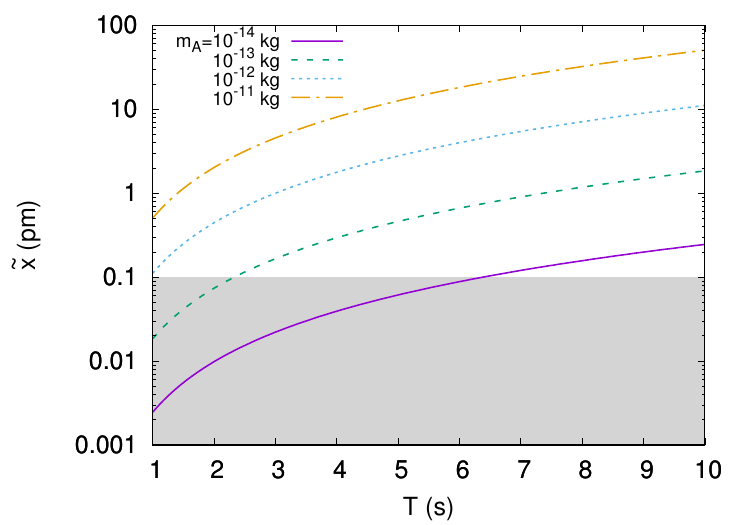}
  \caption{Displacement $\tilde x$ of a silica test mass $m_B = 10^{-14}$
  as a function of the evolution time $T$, 
  for different values of the delocalized source mass $m_A$. 
 The shaded area shows the displacement below the 
current resolution limit (0.1 pm).
 Test and source masses are assumed to be made of silica.}
  \label{fig6}
\end{figure}

This lower bound for the source mass $m_A$ could be reduced for evolution times 
longer than $T = 1 \, {\rm s}$.
This is shown in Fig.~\ref{fig6}, which displays the displacement
$\tilde x$ as a function of the evolution time for different values
of the delocalized source mass $m_A$. Displacement would
exceed the spatial resolution of 0.1 pm and be detectable outside the
shaded area in Fig.~\ref{fig6}. 

As already mentioned,
maintaining coherence on time-scales $T \gtrsim 1 \, {\rm s}$ 
 for masses $m_A \approx 10^{-14}$
is clearly not an easy task.
The system has to be cooled down
to very low temperatures since thermal photons alone are 
able to wash out coherence
very rapidly,  no matter how good 
the vacuum is in the experimental setup \cite{Joos}.
Experimental schemes 
able to reach the above time-scales 
for masses approximately equal to $10^{-14} \, \rm{kg}$
and separations of the order of their size
have been proposed in \cite{Pino}.
They require temperatures of a few tens of
mK at most and pressures at the state-of-the 
art level of $P \approx 10^{-20} \, \rm{bar}$.     
  
%%%%%%%%%%%%%%%%%%%%%%%%%%%%%%%

\section{Conclusion}

In summary,
we have described a possible setup in a laboratory setting
to check for
the quantum nature of the gravitational field.
This was done by exploiting the ability of the gravitational
field to become entangled with the superposed positions
of a single delocalized particle, if the field is quantum in nature.

In a way, our proposal can be regarded 
as an attempt to turn part of the gedanken experiment 
of \cite{Baym, Mari, Belenchia1, Belenchia2, Rydving, Danielson, Pesci}
(where a test particle probes the gravitational field 
of a delocalized particle while this recombines) into something real.
Specifically, it is an attempt for a practical implementation of
the which-path part of the gedanken experiment, consisting in 
the ability of the test particle to discriminate which branch 
is taken in the superposition, if the field requires a quantum description.
  
To turn the thought experiment into something real,
we have shown that the test particle should be placed
very close to the delocalized one, thus potentially clashing 
with Casimir-Polder effects, 
which are expected to overwhelm the signal
we are after.
We discussed a way to overcome this difficulty,
basically using electrical screening through metallic plates,
as conceived also, for different settings, in 
\cite{Westphal, van de Kamp, Yi}.
In the setup we have considered, it turns out that Casimir-Polder forces
can actually be used, instead of simply be fought,
to help in balancing the weight of the particles 
(possibly eliminating altogether the need of optical levitation).

The proposed setup requires a good enough sensitivity
to the displacement of the test particle, 
which, if gravity is quantum, is differential in the two superposed branches.
This involves the use of delocalized particles 
with masses greater than approximately $10^{-14} \, {\rm kg}$
and separations approximately equal to $10-20 \, {\rm \mu m}$, 
which should
keep quantum coherence for time intervals as long as $1-10 \, {\rm s}$.
These requirements are not so different from other proposals
for gravity-induced entanglement experiments \cite{Bose, van de Kamp},
but, as mentioned, remain extremely challenging at present.
Innovative techniques to prepare such large masses 
in a delocalized
state with such large separations have crucially 
to be envisaged \cite{Pino},   
including severe cooling 
of the system ($T \lesssim 10 \, {\rm mK}$)
and extremely low pressures ($P \lesssim 10^{-20} \, {\rm bar}$)
to keep control on its evolution. 
As for the test particle,
the needed spatial resolution 
with optical tweezers through ground-state cooling 
requires masses greater than approximately $10^{-14} \, {\rm kg}$,
two orders of magnitude larger
than the largest mass controlled so far
with these techniques.
Finally, even if our investigation has been only 
at a proof-of-concept level,
we have discussed
that possible
imperfections in the components of the experimental apparatus,
in particular concerning the accuracy of the planarity of the
metallic screens
and random irregularities of the surfaces 
could critically affect the feasibility of the experiment
and deserve careful consideration.
In this respect, we have argued that by keeping 
the lateral scale for the surface roughness within 100 nm
and the amplitude of fluctuations within 10 nm, the 
effects of irregularities could be kept under control.
Ion-blocking layers applied to the surfaces and 
a fine-tuning of the balance between the Casimir-Polder force and
the weight provided by an additional optical levitation system
could finally eliminate also disturbances originating
by electromagnetic patch potentials.

We hope that all of these challenges, 
apparently
more technical than 
fundamental in nature,
 might be successfully 
addressed in future experiments.

\begin{acknowledgements}
We thank Alessio Belenchia, Sumanta Chakraborty, Dawood Kothawala,
 Lorenzo Piroli, 
Massimiliano Rossi, Valerio  Scarani, and Peter Sidajaya for reading the manuscript and/or
discussions. Partial financial support 
from INFN through grants FLAG and QUANTUM is acknowledged.
\end{acknowledgements}

%%%%%%%%%%%%%%%%%%%%%%%%%%%%%%%%%%%%%%%%%%%%%%%%%%

\end{document}